\documentclass[12pt]{article}
\usepackage{graphicx}
\usepackage{subfigure}

\newcommand\pubnumber{}
\newcommand\pubdate{\today}

\def\tohoku{Department of Physics, Tohoku University\\
6-3, Aramaki Aza-Aoba, Aoba-ku, Sendai, Miyagi 980-8578, Japan\\
~\\
JSPS Research Fellow}
\def\Title#1{\begin{center} {\Large #1 } \end{center}}
\def\Author#1{\begin{center}{ \sc #1} \end{center}}
\def\Address#1{\begin{center}{ \it #1} \end{center}}

\newcommand\pubblock{\rightline{\begin{tabular}{l} \pubnumber\\
         \pubdate  \end{tabular}}}
\newenvironment{Abstract}{\begin{quotation}  }{\end{quotation}}
\newenvironment{Presented}{\begin{quotation} \begin{center} 
             \if0 PRESENTED AT \fi \end{center}\bigskip 
      \begin{center}\begin{large}}{\end{large}\end{center} \end{quotation}}

\begin{document}
\begin{titlepage}
\pubblock

\vfill
\Title{Belle time-integrated $\phi_3$ ($\gamma$) measurements}
\vfill
\Author{Yasuyuki Horii\\(Belle Collaboration)}
%\Author{ Despina Reggiano\support}
\Address{\tohoku}
\vfill
\begin{Abstract}
We report recent results by the Belle collaboration on the determination
of the $CP$-violating angle $\phi_3$ ($\gamma$) using time-integrated methods.
\end{Abstract}
\vfill
\begin{Presented}
Proceedings of CKM2010,\\
{\normalsize the 6th International Workshop on the CKM Unitarity Triangle,\\
University of Warwick, UK, 6-10 September 2010}
\end{Presented}
\vfill
\end{titlepage}
\def\thefootnote{\fnsymbol{footnote}}
\setcounter{footnote}{0}

\section{Introduction}

Precise measurements of the parameters of the standard model
are fundamentally important and may reveal new physics.
The Cabibbo-Kobayashi-Maskawa (CKM) matrix~\cite{Cabibbo, KM} consists of
weak-interaction parameters for the quark sector,
and the phase $\phi_3$ (also known as $\gamma$) is defined by the elements of the CKM matrix
as $\phi_3 \equiv \arg{(-V_{ud}{V_{ub}}^*/V_{cd}{V_{cb}}^*)}$.
This phase is less accurately measured than the two other angles
$\phi_1$ ($\beta$) and $\phi_2$ ($\alpha$) of the unitarity triangle.\footnote{
The angles $\phi_1$ and $\phi_2$ are defined as $\phi_1 \equiv \arg{(-V_{cd}{V_{cb}}^*/V_{td}{V_{tb}}^*)}$
and $\phi_2 \equiv \arg{(-V_{td}{V_{tb}}^*/V_{ud}{V_{ub}}^*)}$.}
%In this brief note, we report a few studies by Belle collaboration related to measurement of $\phi_3$.

\begin{figure}[htb]
 \begin{center}
  \leavevmode
  \subfigure
  {\includegraphics[bb=103 350 448 495, width=0.4\textwidth]{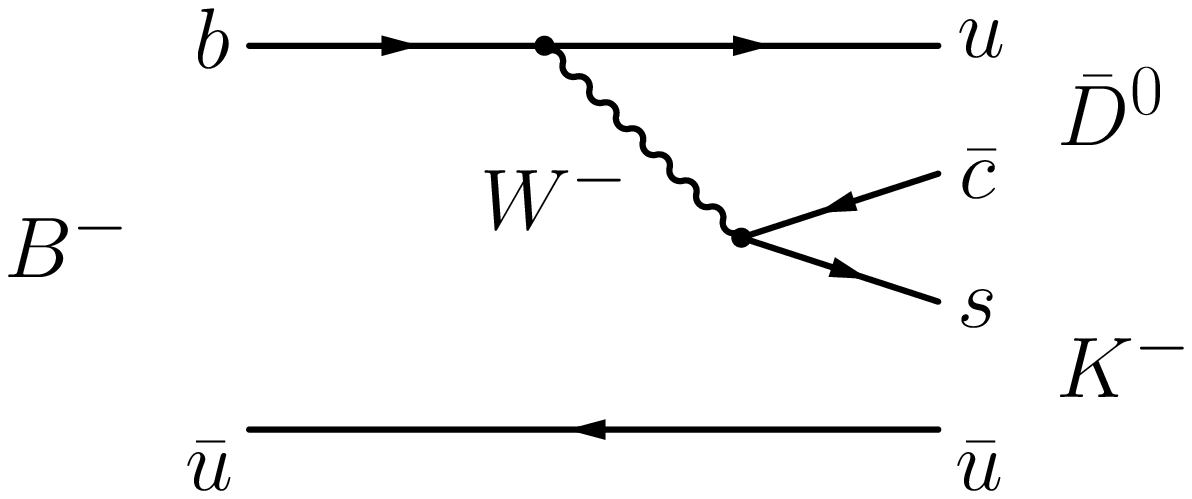}}
  \hspace{8mm}
  \subfigure
  {\includegraphics[bb=103 350 448 495, width=0.4\textwidth]{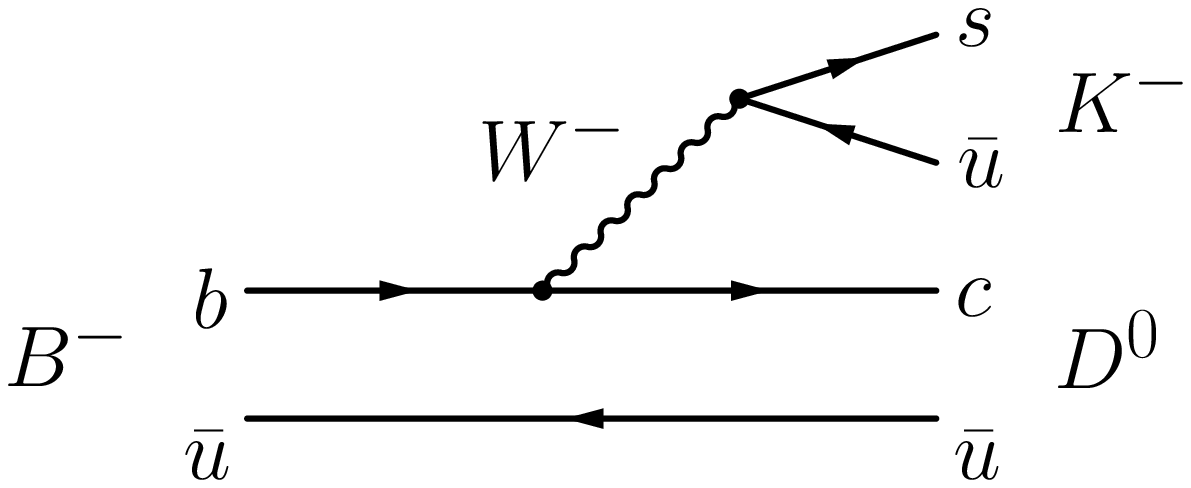}}
  \caption{Diagrams for the $B^- \rightarrow \bar{D}^0 K^-$ and $B^- \rightarrow D^0 K^-$ decays.}
  \label{fig:diagrams}
 \end{center}
\end{figure}

%The possibility of large $CP$ asymmetries in the decays $B \rightarrow DK$ are
%first discussed by I.~Bigi, A.~Carter, and A.~Sanda.\cite{Bigi_Sanda}
%Since then, several methods for measuring $\phi_3$ using $B \rightarrow DK$ decays have been proposed.
In the usual quark phase convention where large complex phases appear only in $V_{ub}$ and $V_{td}$~\cite{Wolfenstein},
the measurement of $\phi_3$ is equivalent to the extraction of the phase of $V_{ub}$
relative to the phases of other CKM matrix elements except for $V_{td}$.
Figure~\ref{fig:diagrams} shows the diagrams for $B^- \rightarrow \bar{D}^0K^-$ ($b\rightarrow u$)
and $B^- \rightarrow D^0K^-$ ($b\rightarrow c$) decays.\footnote{
Charge conjugate modes are implicitly included unless otherwise stated.}
By analyzing the interfering processes produced when $\bar{D}^0$ and $D^0$ decay to the same final states,
we extract $\phi_3$ as well as relevant dynamical parameters.
We define the magnitude of the ratio of amplitudes $r_B = |A(B^-\rightarrow \bar{D}^0K^-)/A(B^-\rightarrow D^0K^-)|$
and the strong phase difference $\delta_B = \delta(B^-\rightarrow \bar{D}^0K^-)-\delta(B^-\rightarrow D^0K^-)$,
which are crucial parameters needed in the extraction of $\phi_3$.
%The feasibility for measuring $\phi_3$ crucially depends on the size of $r_B$,
%which is predicted to be around 0.1-0.2 by taking a product of the ratio of the CKM matrix elements
%$|V_{ub}{V_{cs}}^*/V_{cb}{V_{us}}^*|$ and the color suppression factor.
In this report, we show recent results by the Belle collaboration on the determination of $\phi_3$.

\section{Result for $B^-\rightarrow D^{(*)}K^-,~D\rightarrow K_S\pi^+\pi^-$}

One of most promising ways of measuring $\phi_3$ uses the decay
$B^-\rightarrow DK^-,~D\rightarrow K_S\pi^+\pi^-$~\cite{Dalitz1, Dalitz2},
where $D$ indicates $\bar{D}^{0}$ or $D^{0}$.
The method is based on the fact that the amplitudes for $B^\pm$ can be expressed by
%\vspace{-1mm}
\begin{equation}
 M_\pm = f(m_\pm^2, m_\mp^2) + r_B e^{\pm i\phi_3 + i\delta_B} f(m_\mp^2, m_\pm^2),
\end{equation}
where $m_\pm^2$ are defined as Dalitz plot variables $m_\pm^2 \equiv m_{K_S\pi^\pm}^2$,
and $f(m_+^2, m_-^2)$ is the amplitude of the $\bar{D}^0\rightarrow K_S\pi^+\pi^-$ decay.
By applying a fit on $m_\pm^2$, $\phi_3$ is extracted with $r_B$ and $\delta_B$.
The decay $B^-\rightarrow D^*K^-$ can also be used by reconstructing $D^*$ from $D\pi^0$ or $D\gamma$,
for which the parameters $r_B^*$ and $\delta_B^*$ are introduced.

\begin{figure}[htb]
 \begin{center}
  \leavevmode
  \subfigure
  {\includegraphics[width=0.4\textwidth]{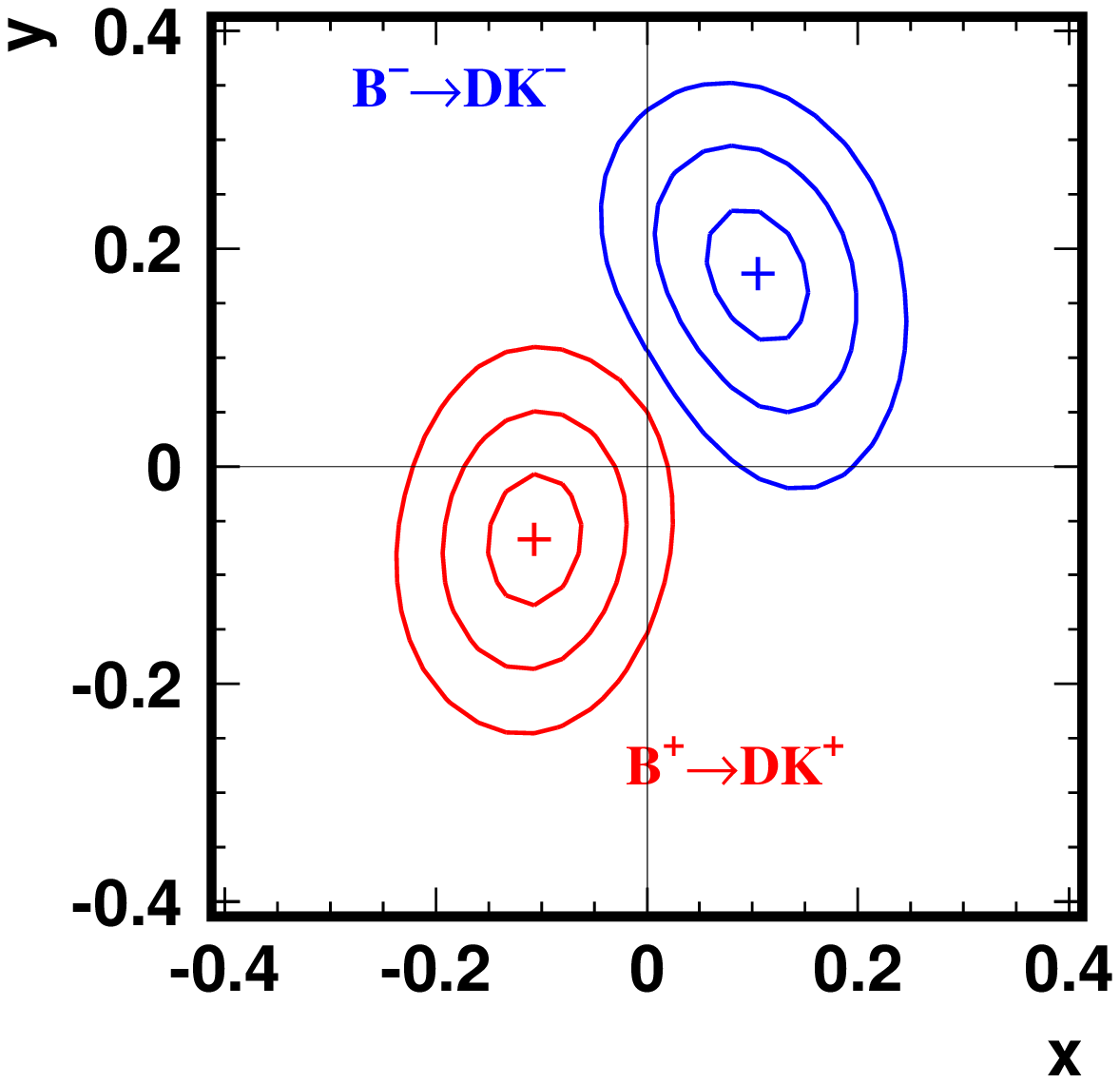}}
  \hspace{6mm}
  \subfigure
  {\includegraphics[width=0.4\textwidth]{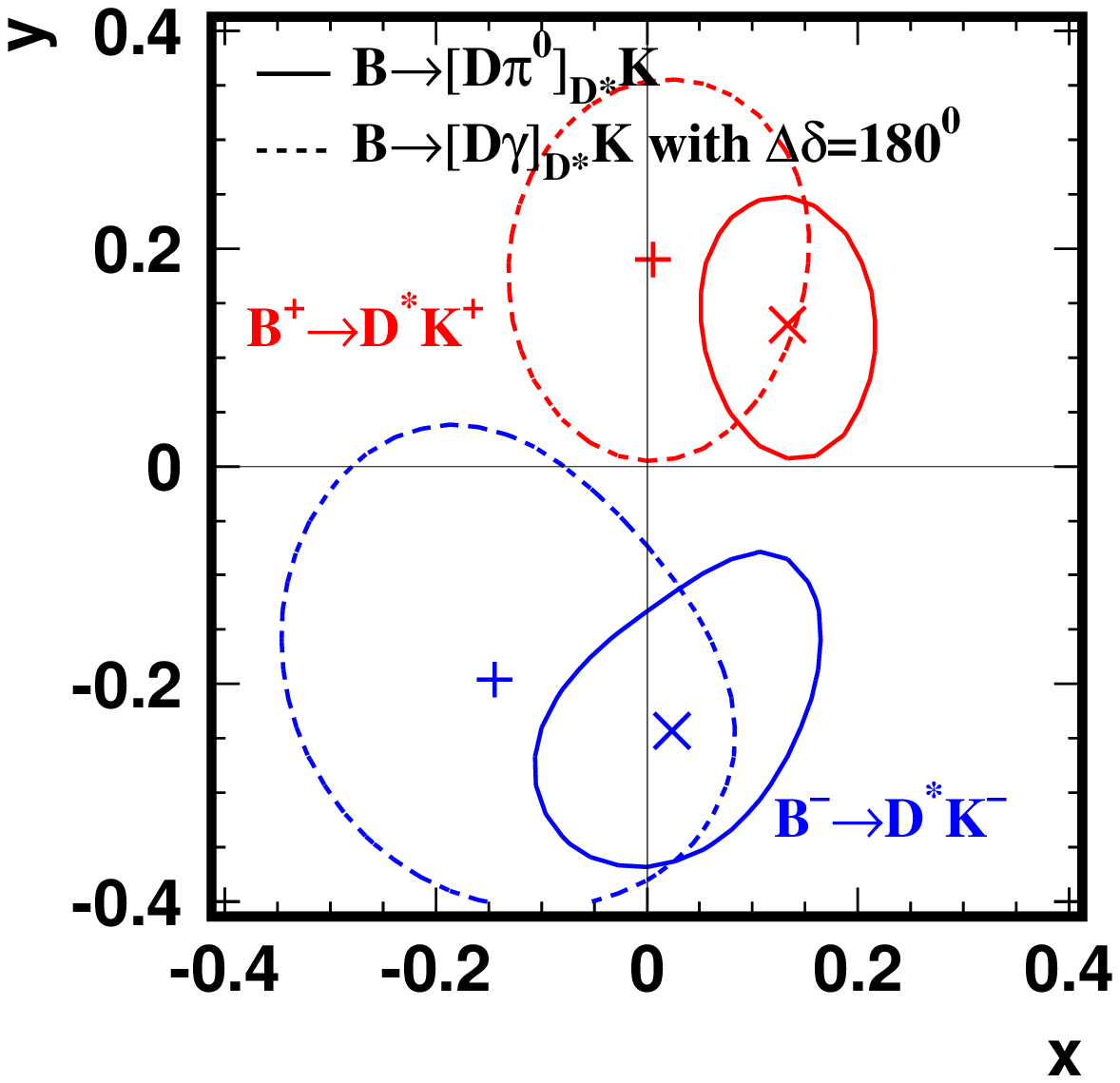}}
  \caption{Results of the fits for $B^-\rightarrow DK^-$ (left) and
  $B^-\rightarrow D^*K^-$ (right) samples,
  where the contours indicate 1, 2, and 3 (left) and 1 (right) standard-deviation regions.}
  \label{fig:dalitz_xy}
 \end{center}
\end{figure}

The result~\cite{Dalitz_result} is based on a data sample
that contains $6.6\times 10^{8}$ $B\bar{B}$ pairs.
The amplitude $f(m_+^2, m_-^2)$ is obtained
by a large sample of $\bar{D}^0\rightarrow K_S\pi^+\pi^-$ decays
produced in continuum $e^+e^-$ annihilation,
where the isobar model is assumed with Breit-Wigner functions for resonances.
%where the flavors are tagged
%by the soft-pion charges in $D^{*\pm}\rightarrow D\pi_{\rm soft}^\pm$.
The background fractions are determined depending on $\Delta E \equiv E_B - E_{\rm beam}$,
$M_{\rm bc} \equiv \sqrt{E_{\rm beam}^2 - |\vec{p}_B|^2}$,
and event-shape variables for suppressing the $e^+e^-\rightarrow q\bar{q}$ ($q=u, d, s, c$) background,
where $E_B$ ($\vec{p}_B$) and $E_{\rm beam}$ are defined in the $e^+e^-$ center-of-mass frame
as the energy (the momentum) of the reconstructed $B$ candidates and the beam energy, respectively.
Using obtained amplitude $f(m_+^2, m_-^2)$ and background fractions, the fit on $m_\pm^2$ is performed
with the parameters $x_\pm = r_\pm \cos{(\pm\phi_3+\delta_B)}$ and $y_\pm = r_\pm \sin{(\pm\phi_3+\delta_B)}$,
where we take $r_B$ separately for $B^\pm$ as $r_\pm$.
The results are shown in Figure~\ref{fig:dalitz_xy} for $B^-\rightarrow DK^-$ and $B^-\rightarrow D^*K^-$.
%The separation for the values of total phases $\phi_3 + \delta_B$ and $-\phi_3 + \delta_B$
%indicate an evidence of the $CP$ violation.
The separations with respect to the charges of $B^\pm$ indicate an evidence of the $CP$ violation.
From the results of the fits, we measure
%\vspace{-1mm}
\begin{equation}
\phi_3 = 78.4^\circ~^{+10.8^\circ}_{-11.6^\circ}({\rm stat}) \pm 3.6^\circ({\rm syst}) \pm 8.9^\circ({\rm model})
\end{equation}
as well as $r_B = 0.161^{~+0.040}_{~-0.038}\pm 0.011^{~+0.050}_{~-0.010}$,
$r_B^* = 0.196^{~+0.073}_{~-0.072}\pm 0.013^{~+0.062}_{~-0.012}$,
%$\delta_B = (137.4^{~+13.0}_{~-15.7}\pm 4.0\pm 22.9)^\circ$, and
%$\delta_B^* = (341.7^{~+18.6}_{~-20.9}\pm 3.2\pm 22.9)^\circ$.
$\delta_B = 137.4^\circ~^{+13.0^\circ}_{-15.7^\circ}\pm 4.0^\circ\pm 22.9^\circ$, and
$\delta_B^* = 341.7^\circ~^{+18.6^\circ}_{-20.9^\circ}\pm 3.2^\circ\pm 22.9^\circ$.
The model error is due to the uncertainty in determining $f(m_+^2, m_-^2)$.
Note that it is possible to eliminate this uncertainty
using constraints obtained by analyzing $\psi(3770)\rightarrow D^0\bar{D}^0$~\cite{Dalitz_MI}.
\if0
\begin{figure}[htb]
 \begin{center}
  \leavevmode
  \subfigure
  {\includegraphics[width=0.28\textwidth]{fig_dalitz_1a.eps}}
  \hspace{2mm}
  \subfigure
  {\includegraphics[width=0.28\textwidth]{fig_dalitz_1c.eps}}
  \hspace{2mm}
  \subfigure
  {\includegraphics[width=0.28\textwidth]{fig_dalitz_1e.eps}}\\
  \subfigure
  {\includegraphics[width=0.28\textwidth]{fig_dalitz_1b.eps}}
  \hspace{2mm}
  \subfigure
  {\includegraphics[width=0.28\textwidth]{fig_dalitz_1d.eps}}
  \hspace{2mm}
  \subfigure
  {\includegraphics[width=0.28\textwidth]{fig_dalitz_1f.eps}}
  \caption{The distributions of $\Delta E$, plotted with $M_{\rm bc} > 5.27~{\rm MeV/}c^2$,
  and $M_{\rm bc}$, plotted with $|\Delta E| < 30~{\rm MeV}$, for $B^-\rightarrow DK^-$ (left),
  $B^-\rightarrow D^*K^-$ with $D^*\rightarrow D\pi^0$ (middle) and
  $B^-\rightarrow D^*K^-$ with $D^*\rightarrow D\gamma$ (right).}
  \label{fig:dalitz_2d}
 \end{center}
\end{figure}
\fi

\section{Result for $B^-\rightarrow DK^-,~D\rightarrow K^+\pi^-$}

The effect of $CP$ violation can be enhanced,
if the final state of the $D$ decay following to the $B^-\rightarrow DK^-$
is chosen so that the interfering amplitudes have comparable magnitudes~\cite{ADS}.
%The archetype uses the $D$ decays to $K^+\pi^-$,
The decay $D\rightarrow K^+\pi^-$ is a particularly useful mode;
%where the color-suppressed $B$ decay followed by the Cabibbo-favored $D$ decay
%interferes with the color-favored $B$ decay followed by the Cabibbo-suppressed $D$ decay.
the usual observables are the partial rate ${\cal R}_{DK}$ and the $CP$-asymmetry ${\cal A}_{DK}$ defined as
%For this mode, usual observables are the partial rate ${\cal R}_{DK}$ and the $CP$-asymmetry ${\cal A}_{DK}$,
%\vspace{-1mm}
\begin{eqnarray}
  {\cal R}_{DK} &\equiv& \frac{{\cal B}(B^-\rightarrow [K^+\pi^-]_DK^-)+{\cal B}(B^+\rightarrow [K^-\pi^+]_DK^+)}
  	{{\cal B}(B^-\rightarrow [K^-\pi^+]_DK^-)+{\cal B}(B^+\rightarrow [K^+\pi^-]_DK^+)} \nonumber \\
               &=& r_B^2 + r_D^2 + 2r_Br_D \cos{(\delta_B+\delta_D)}\cos{\phi_3}, \label{eq:rdk} \\
  {\cal A}_{DK} &\equiv& \frac{{\cal B}(B^-\rightarrow [K^+\pi^-]_DK^-)-{\cal B}(B^+\rightarrow [K^-\pi^+]_DK^+)}
  	{{\cal B}(B^-\rightarrow [K^+\pi^-]_DK^-)+{\cal B}(B^+\rightarrow [K^-\pi^+]_DK^+)} \nonumber \\
               &=& 2r_Br_D\sin{(\delta_B+\delta_D)}\sin{\phi_3}/{\cal R}_{DK}, \label{eq:adk} 
\end{eqnarray}
where $[f]_D$ indicates that the state $f$ originates from a $D$ meson,
$r_D = |A(D^0\rightarrow K^+\pi^-)/A(D^0\rightarrow K^-\pi^+)|$,
and $\delta_D = \delta(D^0\rightarrow K^-\pi^+) - \delta(D^0\rightarrow K^+\pi^-)$.
For the parameters $r_D$ and $\delta_D$, external experimental inputs can be used~\cite{Charm}.

\begin{figure}[htb]
\centering
\includegraphics[width=1.0\textwidth]{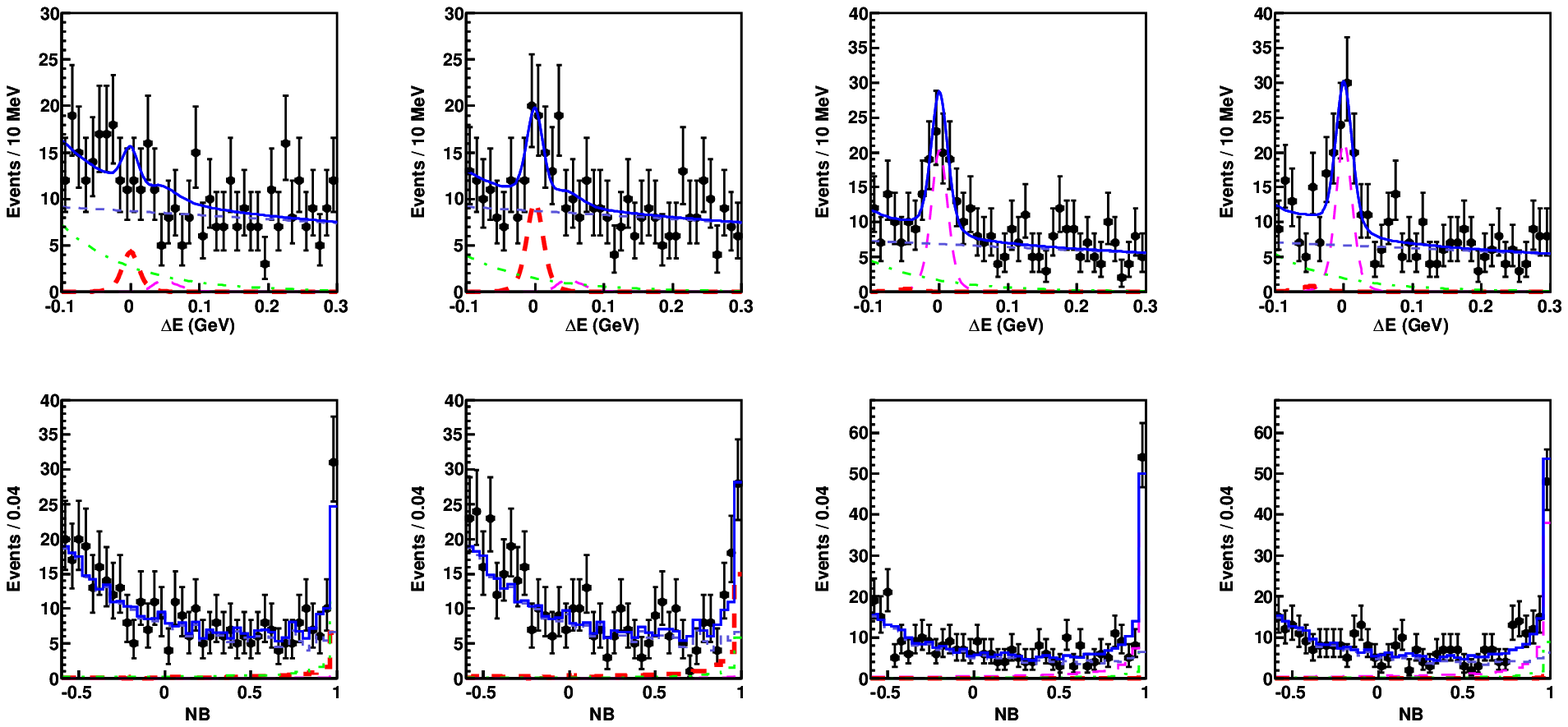}
\caption{The distributions of $\Delta E$ for ${\cal NB} > 0.5$ (top)
and $\cal NB$ for $|\Delta E| < 40~{\rm MeV}$ (bottom)
on the suppressed modes $DK^-$, $DK^+$, $D\pi^-$, and $D\pi^+$ from left to right.
%The fitted sample is shown with dots with error bars and the PDF is shown with the solid blue curve, for which
The components are \if0 shown with \fi thicker long-dashed red ($DK$), thinner long-dashed magenta ($D\pi$),
dash-dotted green ($B\bar{B}$ background), and dashed blue ($q\bar{q}$ background).}
\label{fig:ads}
\end{figure}

In this report, we show a preliminary result %for $B^-\rightarrow D K^-,~D\rightarrow K^+\pi^-$
based on a data sample that contains $7.7\times 10^{8}$ $B\bar{B}$ pairs
(the full data sample collected by Belle at $\Upsilon(4S)$ resonance).
The decay $B^-\rightarrow D \pi^-$ is also analyzed similarly as a reference mode.
%We also similarly analyze $B^-\rightarrow D \pi^-,~D\rightarrow K^+\pi^-$ as a reference mode.
For the largest background from the continuum process $e^+e^-\rightarrow q\bar{q}$,
we apply the new method of the discrimination based on NeuroBayes neural network~\cite{NeuroBayes}.
The inputs are a Fisher discriminant of modified Super-Fox-Wolfram moments,
cosine of the decay angle of $D\rightarrow K^+\pi^-$,
vertex separation between the reconstructed $B$ and the remaining tracks,
and seven other variables. %which specify the difference between the signal and the background.
%By this method, $\sim 2$ times larger significance is expected.
The signal is extracted by a two-dimensional fit on $\Delta E$ and NeuroBayes output ($\cal NB$),
where we simultaneously fit for $DK^-$, $DK^+$, $D\pi^-$, and $D\pi^+$,  as shown in Figure~\ref{fig:ads}.
As a result, we obtain
%\vspace{-1mm}
\begin{eqnarray}
  {\cal R}_{DK} &=& [1.62\pm0.42({\rm stat})^{~+0.16}_{~-0.19}({\rm syst})] \times10^{-2}, \\
  {\cal A}_{DK} &=& -0.39\pm 0.26({\rm stat})^{~+0.06}_{~-0.04}({\rm syst}),\\
  {\cal R}_{D\pi} &=& [3.28\pm0.37({\rm stat})^{~+0.22}_{~-0.23}({\rm syst})] \times10^{-3},\\
  {\cal A}_{D\pi} &=& -0.04\pm 0.11({\rm stat})^{~+0.01}_{~-0.02}({\rm syst}),
\end{eqnarray}
where the first evidence of the suppressed $DK$ signal
is obtained with a significance $3.8\sigma$ including systematic error.
%Our measurements are most precise to date.
%The value of $\phi_3$ will be fitted with other relevant observables.
Our study will make a significant contribution to a model-independent extraction of $\phi_3$
by combining relevant observables, e.g., the partial rates and the $CP$-asymmetries for $D\rightarrow CP$ eigenstates~\cite{GW}.
%The constraint on $\phi_3$ can be obtained by including relevant observables,
%e.g., the partial rates and the $CP$-asymmetries for $D\rightarrow CP$ eigenstates~\cite{GW}.
%$B^-\rightarrow D_{CP}K^-$~\cite{GW}.

%%%%%%%%%%%%%%%%%%%%%%%%%%%%%%%%%%%%%%%%%%%%%%%%%%%%%%%%%%%%%%%%%%%%%%%%%
%%
%%   use this format to include an .eps figure into your paper
%%
%\begin{figure}[htb]
%\centering
%\includegraphics[height=1.5in]{magnet}
%\caption{Plan of the magnet used in the mesmeric studies.}
%\label{fig:magnet}
%\end{figure}
%%%%%%%%%%%%%%%%%%%%%%%%%%%%%%%%%%%%%%%%%%%%%%%%%%%%%%%%%%%%%%%%%%%%%%%%%%%

%%%%%%%%%%%%%%%%%%%%%%%%%%%%%%%%%%%%%%%%%%%%%%%%%%%%%%%%%%%%%%%%%%%%%%%%%
%%
%%   use this format to include a LaTeX table  into your paper
%%
%\begin{table}[thb]
%\begin{center}
%\begin{tabular}{l|ccc}  
%Patient &  Initial level($\mu$g/cc) &  w. Magnet &  
%w. Magnet and Sound \\ \hline
% Guglielmo B.  &   0.12     &     0.10      &     0.001  \\
% Ferrando di N. &  0.15     &     0.11      &  $< 0.0005$ \\ \hline
%\end{tabular}
%\caption{Blood cyanide levels for the two patients.}
%\label{tab:blood}
%\end{center}
%\end{table}
%%%%%%%%%%%%%%%%%%%%%%%%%%%%%%%%%%%%%%%%%%%%%%%%%%%%%%%%%%%%%%%%%%%%%%%%%%%

\section{Conclusion}

In conclusion, recent results on the decays $B^-\rightarrow D^{(*)}K^-$
followed by $D\rightarrow K_S\pi^+\pi^-$ and $D\rightarrow K^+\pi^-$ are reported.
By the Dalitz-plot analysis for $D\rightarrow K_S\pi^+\pi^-$, the value of $\phi_3$ is measured to be
$\phi_3 = 78.4^\circ~^{+10.8^\circ}_{-11.6^\circ}({\rm stat}) \pm 3.6^\circ({\rm syst}) \pm 8.9^\circ({\rm model})$.
For $D\rightarrow K^+\pi^-$, preliminary results on
the partial rate ${\cal R}_{DK}$ and the $CP$-asymmetry ${\cal A}_{DK}$
are reported,
where the first evidence of the signal is obtained with a significance~$3.8\sigma$.
%where the constraint on $\phi_3$ will be applied with other experimental observables.
%Our studies are strongly promoting the determination of $\phi_3$.

%\Acknowledgements
%I am grateful to K. Trabelsi and H. Yamamoto for many advises
%essential to these studies.

\end{document}